\documentclass[prl,twocolumn,showpacs,amsmath,amssymb,superscriptaddress]{revtex4-2}

\usepackage{graphicx}% Include figure files
\usepackage{dcolumn}% Align table columns on decimal point
\usepackage{bm}% bold math
\usepackage{color}

\usepackage{amsmath}
\usepackage{amssymb}
\usepackage{latexsym}
\usepackage{multirow}
\usepackage{xcolor}
\usepackage{hyperref}
 
\usepackage{textcomp}

\begin{document}

\title{Macroscopic Response Diagnoses the Noise Sensitivity of Terminal Outcomes}
\date{\today}
\author{Bo Li}\email{libo312@mails.ucas.ac.cn}
\affiliation{Wuhan City Polytechnic, Hongshan District, Wuhan, Hubei Province 430064, China}
\author{Chaoqian Wang}\email{CqWang814921147@outlook.com}
\affiliation{School of Mathematics and Statistics, Nanjing University of Science and Technology, Nanjing, 210094, China}

\begin{abstract}
Can the terminal macroscopic outcome of a many-body system be inferred from its microscopic initial data without simulating the full trajectory?
Rather than construct such a shortcut, we address a more fundamental question for Gaussian microscopic inputs: can any fixed Wiener--Hermite degree retain a nonvanishing fraction of the variance of the terminal outcome as the system grows?
We consider homogeneous systems with independent Gaussian disorder in which all microscopic coordinates are symmetry-equivalent, the terminal event is monotone in each disorder variable, and a uniform disorder shift is exactly equivalent to a control-field shift with a size-independent conversion factor.
Using forward and inverse Gaussian influence bounds together with a Gaussian Russo formula, we derive a directly measurable criterion that is both necessary and sufficient for noise sensitivity.
Specifically, the correlation between the original and coordinate-perturbed terminal outcomes vanishes asymptotically for every fixed nonzero level of coordinatewise noise if and only if the slope of the outcome probability with respect to the control field at the balanced threshold grows more slowly than the square root of the system volume.
Event-driven simulations of the three-dimensional driven random-field Ising model up to linear size 192 find that both the normalized response and the correlations between perturbed samples decrease overall, consistent with the noise-sensitive regime at finite size.
For spatial Stag-Hunt dynamics with prescribed seeds, the criterion generalizes through an effective number of influential coordinates.
Separately, simulations of a path-dependent best-response game show, over the sizes studied, that the terminal equilibrium can depend on the update schedule while still carrying substantial finite-order predictive information.
\end{abstract}
\maketitle

\noindent{\em Introduction--}
Given microscopic initial data---including an initial configuration and a quenched environment---and an evolution rule, can one infer a system's terminal macroscopic outcome without simulating its full trajectory?
Many physical questions require only such an endpoint: which phase is reached, whether a system-spanning event occurs, or whether macroscopic switching occurs before a prescribed control value.
Before seeking a computational shortcut, one can ask a more basic representational question. 
As the system grows, does the orthogonal projection of the terminal outcome onto any fixed Wiener--Hermite degree of the Gaussian microscopic input retain a nonvanishing fraction of its predictive content?

Determinism alone does not answer this question.
A terminal outcome may be an exact function of the microscopic realization yet become asymptotically decorrelated under any fixed nonzero level of coordinatewise noise, with its predictive content moving to progressively higher Wiener--Hermite orders as the system grows.
Conversely, unresolved dynamical choices can send the same initial state to different equilibria while a stable finite-order component remains predictable.
Thus, trajectory complexity, path dependence, and the order of microscopic information required for terminal prediction are distinct.

Noise sensitivity formalizes the first possibility by comparing a binary macroscopic event in two correlated microscopic worlds and asking whether their outcome correlation vanishes with system size \cite{BenjaminiKalaiSchramm1999,SchrammSteif2010,GarbanPeteSchramm2010}.
For symmetric monotone events, sharp-threshold theory connects distributed microscopic influence to collective transitions that sharpen with system size \cite{FriedgutKalai1996}.
For Gaussian inputs, forward and inverse influence bounds relate outcome correlations to the sum of squared coordinate influences \cite{KellerMosselSen2012,KellerMosselSen2014}.
However, individual influences are generally not available as directly measurable bulk observables.
The practical question is whether noise sensitivity can instead be diagnosed from a macroscopic response.

Here we answer this question for balanced, coordinatewise monotone terminal events defined on independent Gaussian disorder and invariant under a permutation group acting transitively on the microscopic coordinates.
We also assume that a uniform disorder shift is exactly equivalent to a control-field shift with a size-independent positive conversion factor.
By combining established forward and inverse Gaussian influence bounds with a Gaussian Russo formula, we eliminate the individual influences in favor of the slope of the macroscopic switching probability.
The terminal event is noise sensitive if and only if the slope of the macroscopic switching probability with respect to the control field at the balanced threshold grows more slowly than the square root of the system volume.
The result converts an abstract microscopic-influence criterion into a measurable test of whether every fixed Wiener--Hermite degree carries an asymptotically vanishing fraction of the variance.

We test the criterion in the quasistatically driven zero-temperature random-field Ising model (RFIM), a canonical model of disorder-induced hysteresis and avalanche criticality \cite{SethnaEtAl1993,DahmenSethna1996,PerkovicDahmenSethna1999}.
Sensitivity of equilibrium RFIM ground states to quenched-disorder perturbations has been studied previously \cite{AlavaRieger1998}; here we instead consider a nonequilibrium hysteretic terminal event and ask what order of microscopic information is required to predict it.
We then use spatial Stag-Hunt dynamics to show how the criterion generalizes when prescribed initial seed configurations break coordinate transitivity \cite{PachecoEtAl2009,Watts2002}.
Finally, a separate path-dependent potential game illustrates a distinct departure from the transitive setting: microscopic noise sensitivity must be distinguished from update-order uncertainty \cite{Blume1995,MondererShapley1996}.

\noindent{\em Terminal-outcome information--}
Consider $N=L^d$ independent standard Gaussian quenched variables $\bm Z=(Z_1,\ldots,Z_N)$, and let $F_L(\bm Z;H)\in\{0,1\}$ be the indicator of a terminal macroscopic event---for example, whether the system has switched phase before the control field reaches $H$.
For each system size $L$, assume that there exists a control value $H_L^\star$---the balanced threshold, also referred to as the median switching point---at which the event is equally likely---i.e., $\mathbb P[F_L(\bm Z;H_L^\star)=1]=1/2$---and denote $F_L(\bm Z)\equiv F_L(\bm Z;H_L^\star)$.
This places the system at its median switching point, where the variance of the binary outcome is maximal.
The centered indicator admits the orthogonal Wiener--Hermite decomposition:
\begin{equation}
	F_L-\frac12=\sum_{m\ge1} f_{L,m},
	\qquad
	p_m(L)=\frac{\lVert f_{L,m}\rVert_2^2}{\operatorname{Var}(F_L)},
	\label{eq:hermite}
\end{equation}
where $f_{L,m}$ is the homogeneous chaos component of total Hermite degree $m$.
Consequently, $p_m(L)\ge0$, $\sum_{m\ge1}p_m(L)=1$, and $p_m(L)$ is the fraction of the outcome variance carried by degree $m$.

Now introduce a correlated disorder realization, $\bm Z^{(c)}=c\bm Z+\sqrt{1-c^2}\,\bm\Xi$, where $0<c<1$ and $\bm\Xi$ is an independent standard Gaussian vector. 
This construction gives a perturbation that preserves the marginal distribution of each coordinate while maintaining a coordinatewise correlation $c$ with the original.
The correlation between the outcomes in the original and perturbed worlds then follows exactly from the Hermite weights:
\begin{equation}
	\mathcal S_L(c)
	=\frac{\operatorname{Cov}[F_L(\bm Z),F_L(\bm Z^{(c)})]}
	{\operatorname{Var}(F_L)}
	=\sum_{m\ge1}p_m(L)\,c^m.
	\label{eq:noiseidentity}
\end{equation}
Consequently, $\mathcal S_L(c)\to0$ for every fixed $c\in(0,1)$ if and only if
\[
\sum_{m=1}^{D}p_m(L)\to0
\qquad
\text{for every fixed }D.
\]
Equivalently, the minimum Hermite degree needed to capture any fixed positive fraction of the outcome variance diverges with $L$.
In this precise sense, the outcome variance escapes every fixed Hermite order as the system grows \cite{KindlerKirshnerODonnell2018, Supplement}.

\noindent{\em Macroscopic criterion--}
Assume that $F_L$ is coordinatewise increasing---larger microscopic values only favor the event---and that a uniform shift of all microscopic coordinates by $t$ is exactly equivalent to shifting the control field by $Rt$, with a conversion factor $R>0$ independent of $L$:
$F_L(\bm Z+t\bm 1;H)=F_L(\bm Z;H+Rt)$.
Let $I_{i,L}^G$ denote the Gaussian geometric influence of coordinate $i$, defined through the Gaussian boundary measure of the event along that coordinate \cite{KellerMosselSen2012,KellerMosselSen2014}.
Let $s_L=\partial_H\mathbb P[F_L(\bm Z;H)=1]|_{H=H_L^\star}$ be the slope of the macroscopic switching probability at the balanced threshold $H_L^\star$.
The Gaussian Russo formula relates the sum of the coordinate influences to the bulk slope; if the event is invariant under a permutation group acting transitively on the microscopic coordinates, all coordinate influences are equal, and hence \cite{KellerMosselSen2012,Supplement}
\begin{equation}
	\sum_{i=1}^{N}I_{i,L}^G=Rs_L,
	\qquad
	\mathcal I_L^{(2)}
	\equiv\sum_{i=1}^{N}(I_{i,L}^G)^2
	=\frac{R^2s_L^2}{L^d}.
	\label{eq:influences}
\end{equation}
Two established Gaussian influence bounds for increasing events now sandwich the normalized outcome correlation: the reverse correlation bound provides a lower bound on $\mathcal S_L(c)$ in terms of $\mathcal I_L^{(2)}$, while the Gaussian BKS bound provides the corresponding upper control \cite{KellerMosselSen2014,Supplement}. 
Thus, for every fixed $c\in(0,1)$,
\begin{equation}
	4c^2\mathcal I_L^{(2)}
	\le \mathcal S_L(c)
	\le C_1[\mathcal I_L^{(2)}]^{C_2(1-c^2)},
	\label{eq:kmsbounds}
\end{equation}
where $C_1,C_2>0$ are independent of $L$, with $C_1$ absorbing the normalization by $\operatorname{Var}(F_L)=1/4$.
Using $\mathcal I_L^{(2)}=R^2s_L^2/L^d$ in the two bounds, the upper bound forces $\mathcal S_L(c)\to0$ when $s_L/L^{d/2}\to0$, while the lower bound shows that $\mathcal S_L(c)\to0$ is impossible unless $s_L/L^{d/2}\to0$. Since $R$ is fixed and nonzero, the constants do not alter the asymptotic threshold. Hence
\begin{equation}
	\boxed{
		\begin{gathered}
			\mathcal S_L(c)\to0
			\quad\text{for every fixed }c\in(0,1)
			\\[-2pt]
			\Longleftrightarrow
			\quad
			s_L/L^{d/2}\to0 .
	\end{gathered}}
	\label{eq:criterion}
\end{equation}
This is the central operational result: within the stated class, a single bulk observable---the slope of the switching probability at the balanced threshold---provides both a necessary and a sufficient diagnostic of terminal-outcome noise sensitivity.
Via the Hermite correlation identity, this criterion simultaneously answers the representational question raised above: $s_L/L^{d/2}\to0$ is exactly the condition under which every fixed Wiener--Hermite degree carries a vanishing fraction of the outcome variance.
If the response scales as $s_L=L^{y_F+o(1)}$ with $y_F<d/2$, the quantitative upper bound implies that the Hermite degree required to capture any fixed fraction of the outcome variance grows at least logarithmically with $L$. 
Thus the outcome variance moves to progressively higher Hermite orders, although this lower bound does not exclude faster growth \cite{Supplement}.

\noindent{\em Driven-RFIM test--}
We consider the three-dimensional RFIM
\begin{equation}
	\mathcal H=-\sum_{\langle ij\rangle}S_iS_j
	-\sum_i(H+RZ_i)S_i,
	\qquad S_i\in\{-1,1\},
	\label{eq:rfim}
\end{equation}
on a periodic cubic lattice of linear size $L$.
Starting with the system in negative saturation, we increase $H$ quasistatically and fully relax each avalanche before the next field increment.
Ferromagnetic no-passing and Abelian relaxation ensure that, at each $H$, the relaxed metastable state is independent of the order in which unstable spins are updated and is therefore a well-defined function of $\bm Z$ and $H$ \cite{SethnaEtAl1993,DharShuklaSethna1997}.

For each realization, let $H_s(\bm Z)$ be the first field at which the relaxed magnetization exceeds $M_\star=0.9$, and define
\begin{equation}
	F_L(\bm Z;H)=\bm 1\{H_s(\bm Z)\le H\}.
	\label{eq:rfimevent}
\end{equation}
Increasing any $Z_i$ cannot delay an upward spin flip; a uniform shift of all $Z_i$ is exactly equivalent to $H\mapsto H+Rt$; and lattice translations act transitively on the coordinates.
The RFIM therefore satisfies every hypothesis of Eq.~\eqref{eq:criterion}.

We use independent calibration ensembles to determine balanced-threshold fields and event-driven avalanche dynamics to simulate systems with $R=2.16$ for $L=8$ to $192$ \cite{KuntzEtAl1999,Supplement}.
For the slope analysis, the largest three sizes, $L=128,160,192$, contain $1000$, $600$, and $450$ independent disorder realizations, respectively.

\begin{figure}[t]
	\includegraphics[width=\columnwidth]{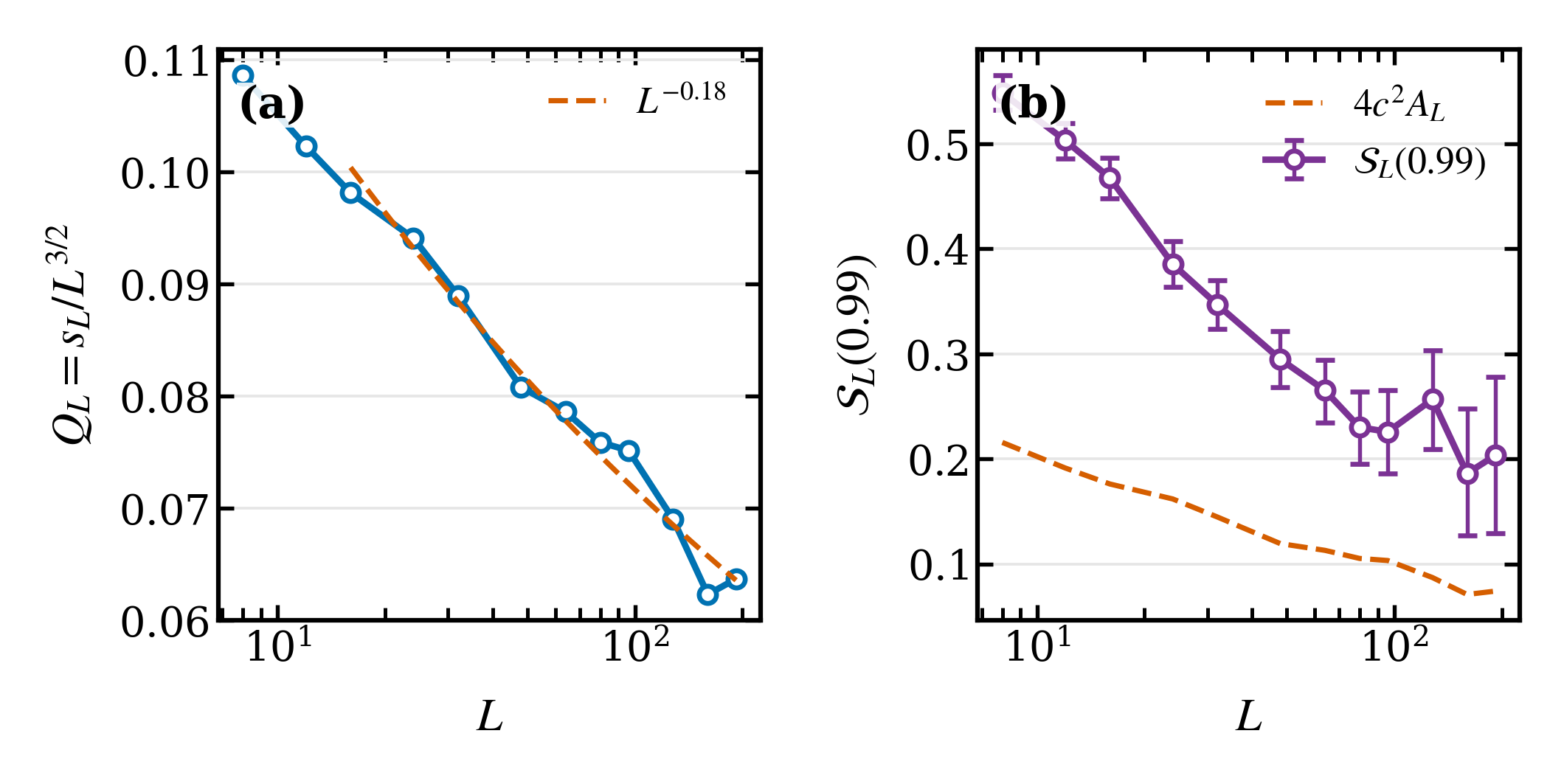}
	\caption{
		Macroscopic diagnostic and microscopic response in the driven three-dimensional RFIM.
		(a) Normalized threshold slope $Q_L=s_L/L^{3/2}$ at $R=2.16$; the dashed line is a finite-size power-law guide with exponent $\widehat y_F-3/2=-0.184$.
		(b) Normalized outcome correlation for disorder pairs with coordinatewise correlation $c=0.99$.
		The dashed lower curve is the plug-in inverse-BKS lower bound $4c^2R^2Q_L^2$, evaluated using the point estimates of $Q_L$ from panel (a).
		Error bars are bootstrap $95\%$ confidence intervals.
		The data are consistent with the thermodynamic limits $Q_L\to0$ and $\mathcal S_L(c)\to0$, though they do not establish these limits definitively.
	}
	\label{fig:criterion}
\end{figure}

Figure~\ref{fig:criterion}(a) tests the macroscopic side of the criterion.
A log--log fit to kernel-density estimates of the switching-field density at the balanced threshold over $L=16$--$192$ gives $\widehat y_F=1.316$ with bootstrap $95\%$ confidence interval $[1.289,1.348]$.
Alternative inverse-IQR and central-window estimators give bootstrap intervals that also lie below $d/2=3/2$ \cite{Supplement}.
Without relying on an exponent fit, the directly estimated $Q_L$ falls overall from approximately $0.098$ at $L=16$ to $0.064$ at $L=192$.

Figure~\ref{fig:criterion}(b) independently probes the microscopic side.
At $c=0.99$, $\mathcal S_L(c)$ falls overall from approximately $0.55$ at $L=8$ to $0.20$ at $L=192$.
The $L=192$ bootstrap interval is $[0.129,0.278]$, while the response-derived plug-in inverse-BKS lower bound is approximately $0.074$.
Because $c$ is close to unity, the available sizes do not prove that the correlation vanishes; they provide finite-size evidence consistent with the same branch as the normalized response.
At the off-critical disorder strength $R=3.0$, by contrast, the inverse interquartile switching-field width normalized by $L^{3/2}$ is approximately size independent over the largest sizes studied \cite{Supplement}.

\begin{figure}[t]
	\includegraphics[width=\columnwidth]{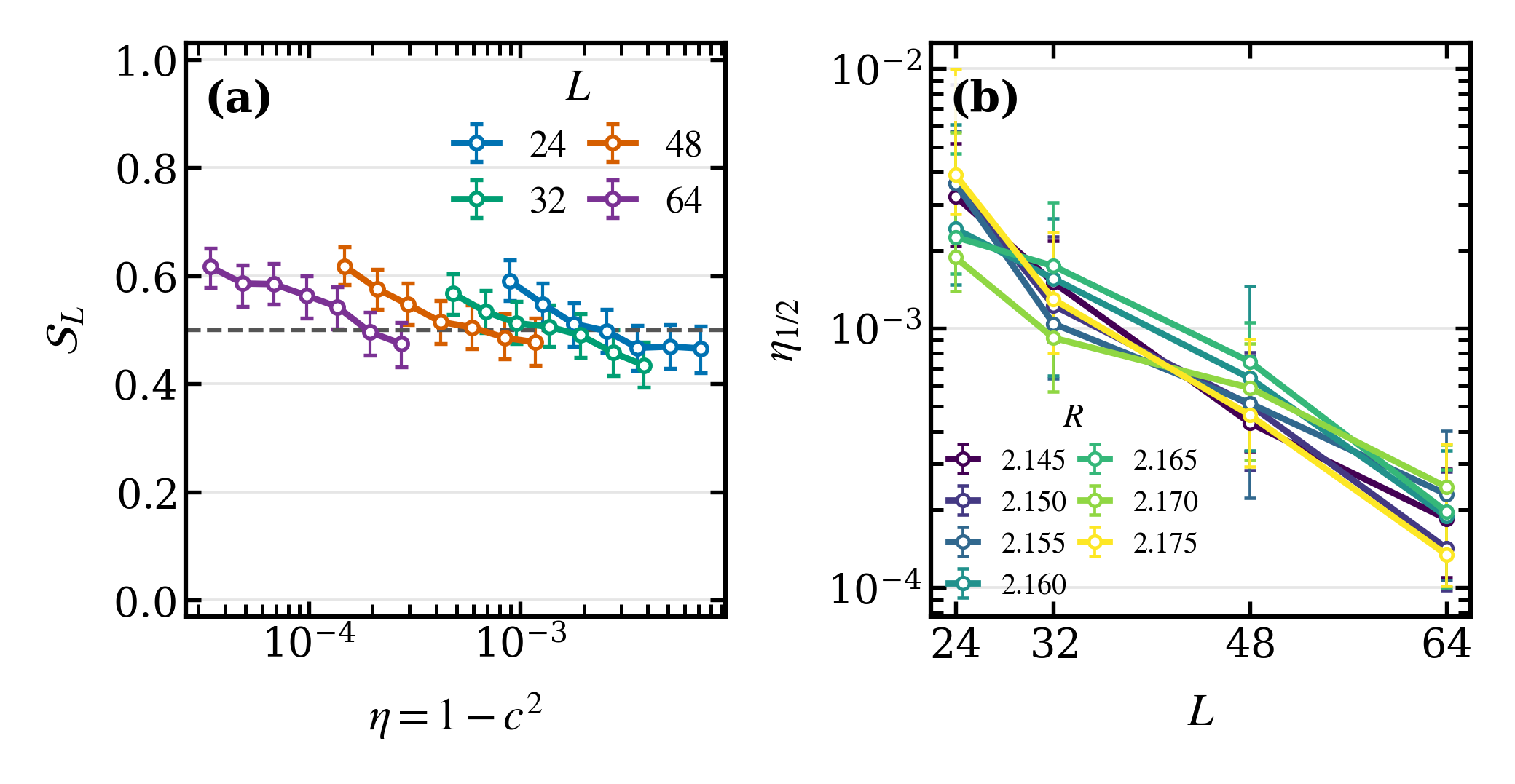}
	\caption{
		Finite-size terminal-outcome noise crossover.
		(a) Raw outcome-correlation curves at $R=2.16$ shift toward smaller noise variance $\eta=1-c^2$ as $L$ increases.
		(b) Half-correlation noise scales $\eta_{1/2}$ for seven disorder strengths; error bars are complete-row bootstrap $95\%$ confidence intervals.
		The narrowing supports noise sensitivity over the simulated range, but we do not infer a universal crossover exponent or a critical disorder value.
	}
	\label{fig:crossover}
\end{figure}

To probe the crossover beyond one value of $c$, we performed an independent scan at $L=24,32,48,64$ and seven disorder strengths $R=2.145$--$2.175$.
All $28$ conditions satisfy prespecified bracketing, balance, shared-world, and independence checks, and every bootstrap draw yields a finite half-correlation point \cite{Supplement}.
As shown in Fig.~\ref{fig:crossover}, the raw crossover narrows with $L$. 
However, a prospectively specified two-variable scaling surface fails the model-selection and stability criteria.
We therefore make no claim for a universal crossover exponent, $R_c$, or $\nu$.

\noindent{\em Scope beyond magnetic dynamics--}
The criterion extends beyond magnetism to a monotone spatial Stag-Hunt adoption process.
For an agent on a lattice of coordination number $z$ with $k_i$ cooperative neighbors, the cooperation payoff advantage is
\begin{equation}
	\Delta u_i=H+\sigma Z_i+k_i(r-t)+(z-k_i)(s-p)
	+\kappa\frac{\binom{k_i}{2}}{\binom z2},
	\label{eq:stag}
\end{equation}
where $r>t\ge p>s$, $\sigma>0$, and $\kappa\ge0$ \cite{PachecoEtAl2009}.
Starting from prescribed committed seeds, $H$ is increased quasistatically, and every strict defector-to-cooperator best response is retained.
Strategic complementarity ensures that every fair update order yields the same monotone closure \cite{Watts2002,Supplement}.

Spatially inhomogeneous prescribed seeds generally break coordinate transitivity, so the geometric influences need not be equal.
For the balanced event that the terminal cooperative fraction reaches a prescribed level $q$, we introduce the effective number of influential preference coordinates,
\begin{equation}
	N_{\mathrm{eff},L,q}
	=\frac{\left(\sum_i I_{i,L,q}^G\right)^2}{\sum_i (I_{i,L,q}^G)^2}.
\end{equation}
The Russo identity then gives the same noise-sensitivity criterion with $L^{d/2}$ replaced by $\sqrt{N_{\mathrm{eff},L,q}}$:
\begin{equation}
	\begin{gathered}
		\mathcal S_{L,q}(c)\to0
		\quad\text{for every fixed }c\in(0,1)
		\\
		\Longleftrightarrow\quad
		\frac{\sigma s_{L,q}}
		{\sqrt{N_{\mathrm{eff},L,q}}}\to0 .
	\end{gathered}
	\label{eq:gamecriterion}
\end{equation}
Thus coordinate transitivity is what reduces the exact criterion to a global-slope test; with arbitrary seeds, one additionally needs the concentration of the site-resolved response.
For $\kappa=0$, the free-agent dynamics maps sample by sample onto the increasing-field zero-temperature RFIM, with the committed seeds represented as pinned up spins; $\kappa>0$ introduces a genuine higher-order assurance interaction \cite{ZimmaroEtAl2024,Supplement}.

\begin{figure}[t]
	\includegraphics[width=\columnwidth]{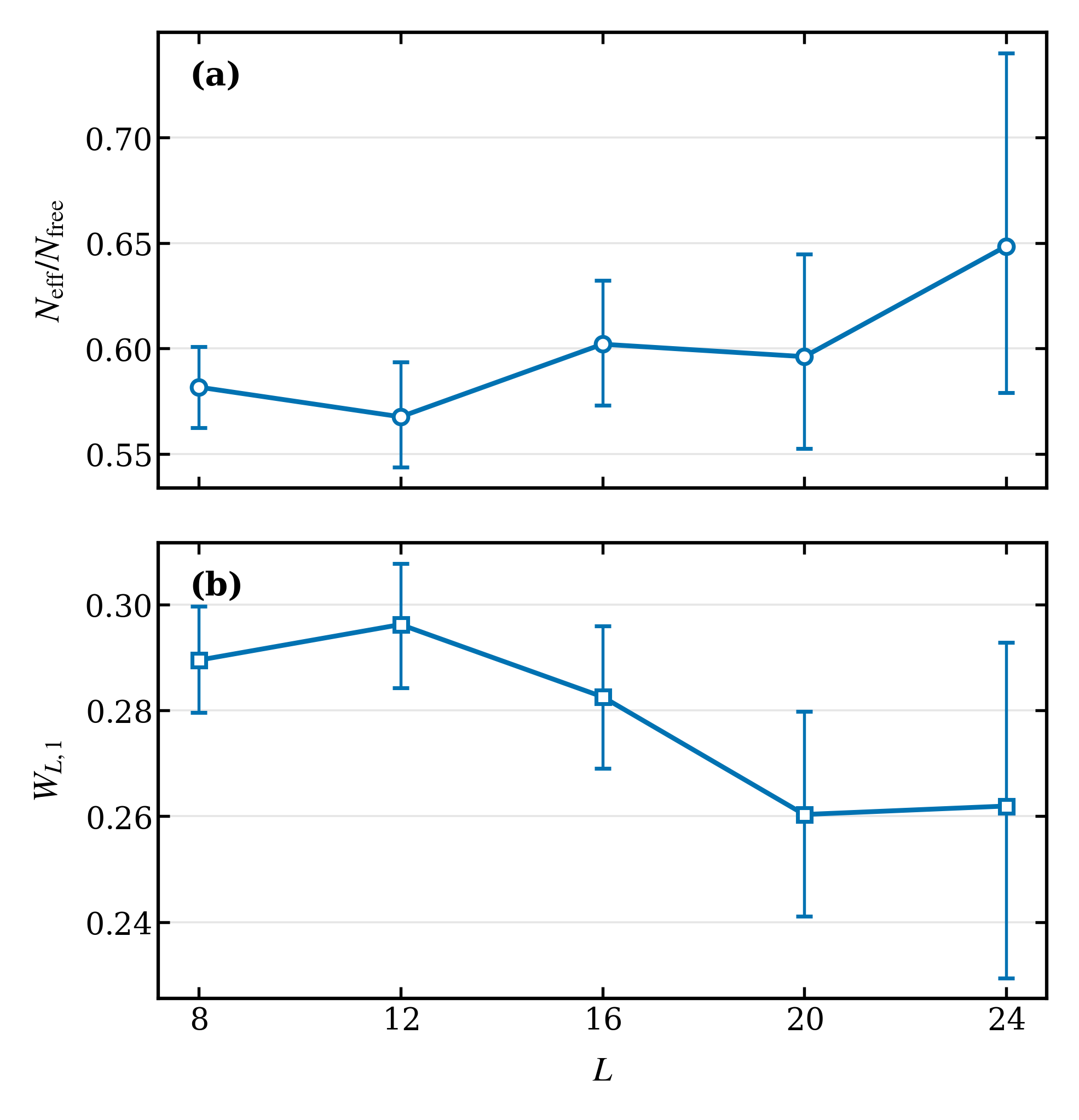}
	\caption{
		Direct response measurement for the nontransitive, plane-seeded Stag-Hunt event.
		(a) Effective influence fraction $N_{\mathrm{eff}}/N_{\mathrm{free}}$.
		(b) Normalized first-order Hermite weight $W_{L,1}$.
		Error bars are two-block chunk-bootstrap $95\%$ intervals; the first-order intervals are obtained by normalizing the squared-influence intervals by the measured Bernoulli variance.
	}
	\label{fig:seededresponse}
\end{figure}

We measure this nontransitive response directly for $(\sigma,\kappa,q)=(1.45,2,0.25)$ with one committed lattice plane, leaving $N_{\mathrm{free}}=L^3-L^2$ uncommitted agents.
Using two independent score-estimation blocks at each size, we find that $N_{\mathrm{eff}}/N_{\mathrm{free}}$ remains of order unity, ranging from $0.568$ to $0.649$ over $L=8,12,16,20,24$.
The corresponding normalized first-order Hermite weight $W_{L,1}$ ranges from $0.260$ to $0.296$.
As shown in Fig.~\ref{fig:seededresponse}, the effective influence fraction shows no systematic decrease over the simulated sizes, while the response spreads progressively beyond the layers adjacent to the seed plane \cite{Supplement}.
These measurements provide finite-size evidence that the response broadens away from the seed plane; no thermodynamic-limit claim is made.

The adoption-only Stag-Hunt process above has an order-independent monotone closure and therefore removes update-order dependence by construction.
To complement this, we study a distinct bidirectional asynchronous best-response game with an exact potential, in which different random strict-improvement schedules can reach different Nash equilibria from the same initial state \cite{Blume1995,MondererShapley1996,Supplement}.
For $L=8$ to $48$, the fraction of binary-outcome variance attributable to the initial state remains between $0.554$ and $0.572$, while approximately $67\%$ of the initial states at the two largest sizes produce both outcomes across the $20$ sampled schedules.
At both $L=32$ and $48$, a mixed Hoeffding--Hermite analysis gives curvewise-bootstrap lower bounds of at least $0.616$ for the fraction of initial-state outcome variance contained through order $10$ and $0.708$ for the fraction contained through order $20$.
These finite-size results show that schedule-dependent equilibria can coexist with a substantial finite-order predictive component; they are not intended as thermodynamic-limit claims.

\noindent{\em Discussion--}
The initial-to-terminal prediction problem separates two questions: whether the terminal outcome admits a stable fixed-order Wiener--Hermite representation, and whether this representation can be computed efficiently without following the trajectory.
Here we answer the first question for a precisely defined class of terminal events driven by independent Gaussian disorder.
Within this class, the volume-normalized response at the balanced threshold vanishes if and only if the terminal outcome is noise sensitive; equivalently, the outcome variance fails to concentrate in any fixed Wiener--Hermite degree.
The RFIM simulations probe both the macroscopic-response and paired-world sides of this equivalence and consistently support the noise-sensitive branch near the disorder-induced critical region, without claiming to establish thermodynamic-limit behavior.

The evolutionary-game realizations illustrate both the scope and the limitations of the criterion.
When prescribed seeds break coordinate transitivity, the system volume is replaced by an effective number of influential coordinates; stochastic update order, by contrast, constitutes a distinct source of uncertainty that need not destroy finite-order predictability.
In the plane-seeded monotone game, the measured effective count remains an $O(1)$ fraction of the free coordinates, while the first-order Hermite weight remains substantial over $L=8$--$24$, providing a finite-size nontransitive counterpart to the transitive criterion.
The criterion yields neither a trajectory-free algorithm for individual realizations nor a worst-case computational-complexity lower bound.
A natural next step is to extend the response-based diagnosis to correlated or non-Gaussian disorder and to dynamics with stochastic transition rules.

\noindent{\em Acknowledgments--}
This work was supported by the National Natural Science Foundation of China (Grant No. 71932008) and the Wuhan City Polytechnic Research Project (Grant No. 2025WHCPB02).
DeepSeek (DeepSeek-V3) was used to assist with analytical derivations, code writing, and draft preparation.
The authors directed its use, independently verified all derivations, code, and numerical results, and take full responsibility for the scientific content.

\clearpage
\onecolumngrid

% Supplemental Material numbering for the combined source file.
% Do not use \appendix here: REVTeX would override the section format.
\setcounter{secnumdepth}{2}
\setcounter{section}{0}
\setcounter{subsection}{0}
\renewcommand{\thesection}{\Roman{section}}
\renewcommand{\thesubsection}{\thesection.\Alph{subsection}}

\renewcommand{\theequation}{S\arabic{equation}}
\setcounter{equation}{0}
\renewcommand{\thefigure}{S\arabic{figure}}
\setcounter{figure}{0}
\renewcommand{\thetable}{S\arabic{table}}
\setcounter{table}{0}

% Distinct PDF anchors after resetting the Supplemental Material counters.
\renewcommand{\theHsection}{supp.\arabic{section}}
\renewcommand{\theHsubsection}{supp.\arabic{section}.\arabic{subsection}}
\renewcommand{\theHequation}{supp.\arabic{equation}}
\renewcommand{\theHfigure}{supp.\arabic{figure}}
\renewcommand{\theHtable}{supp.\arabic{table}}

\section*{Supplemental Material}
This Supplemental Material provides additional theoretical derivations, numerical methods, validation procedures, and finite-size results supporting the Letter.

{\em Section I.---} We establish the relation between terminal-outcome noise sensitivity and fixed-order Wiener--Hermite information, and derive the macroscopic criterion from Gaussian influence bounds and a Russo formula.

{\em Section II.---} We describe the driven random-field Ising model (RFIM), its event-driven implementation, the production samples, and the response-exponent estimates.

{\em Section III.---} We present the RFIM robustness and paired-world analyses, including alternative switching thresholds, multiple noise levels, the confirmatory crossover study, and an off-critical control.

{\em Section IV.---} We give the evolutionary-game extensions, including prescribed-seed Stag-Hunt dynamics, direct measurement of the nontransitive response, and a path-dependent best-response game.

{\em Section V.---} We derive the Bethe-lattice two-world benchmark.

{\em Section VI.---} We clarify the relation between the present terminal-event problem and previous RFIM disorder-chaos studies.

\section{Noise correlation and macroscopic criterion}

\subsection{Noise correlation and the escape of fixed-order information}

Let
\begin{equation}
	f_L(\bm Z)=F_L(\bm Z)-\mathbb E F_L
\end{equation}
and decompose it into homogeneous Wiener--Hermite chaoses,
\begin{equation}
	f_L=\sum_{m\ge1}f_{L,m}.
\end{equation}
Orthogonality gives
\begin{equation}
	\operatorname{Var}(F_L)=\sum_{m\ge1}\|f_{L,m}\|_2^2.
\end{equation}
Define
\begin{equation}
	p_m(L)=
	\frac{\|f_{L,m}\|_2^2}{\operatorname{Var}(F_L)}.
\end{equation}
Then $p_m(L)\ge0$ and $\sum_{m\ge1}p_m(L)=1$, so the $p_m(L)$ form a probability distribution over Hermite orders.
The Ornstein--Uhlenbeck operator $T_c$ satisfies
\begin{equation}
	T_cf_{L,m}=c^m f_{L,m}.
\end{equation}
For
\begin{equation}
	\bm Z^{(c)}=c\bm Z+\sqrt{1-c^2}\,\bm\Xi,
\end{equation}
where $0<c<1$ and $\bm\Xi$ is an independent standard Gaussian vector, we therefore obtain
\begin{align}
	\mathcal S_L(c)
	&=
	\frac{\mathbb E[f_L(\bm Z)f_L(\bm Z^{(c)})]}
	{\operatorname{Var}(F_L)}
	\\
	&=
	\sum_{m\ge1}p_m(L)c^m.
	\label{eq:SidentityS}
\end{align}

For $0<c<1$ and $m\le D$, we have $c^m\ge c^D$. Hence
\begin{equation}
	\mathcal S_L(c)
	=
	\sum_{m\ge1}p_m(L)c^m
	\ge
	c^D\sum_{m\le D}p_m(L).
\end{equation}
Thus $\mathcal S_L(c)\to0$ for any fixed $c\in(0,1)$ implies
\begin{equation}
	\sum_{m\le D}p_m(L)\to0
	\qquad
	\forall D<\infty.
\end{equation}
Intuitively, if all outcome variance resides beyond any fixed finite order, then perturbing the input destroys the correlation.
Consequently the minimum order
\begin{equation}
	D_\varepsilon(L)
	=
	\min\left\{
	D\in\mathbb N_0:
	\sum_{m>D}p_m(L)\le\varepsilon
	\right\}
\end{equation}
diverges for every fixed $0<\varepsilon<1$.
That is, as the system grows, the required order to capture any fixed fraction of the variance grows without bound.

Conversely, if $\sum_{m\le D}p_m(L)\to0$ for every fixed $D$, then, since $\sum_{m>D}p_m(L)c^m\le c^{D+1}$,
\begin{equation}
	\mathcal S_L(c)
	\le
	\sum_{m\le D}p_m(L)+c^{D+1}.
\end{equation}
Taking $L\to\infty$ and then $D\to\infty$ gives $\mathcal S_L(c)\to0$.
The order of limits matters: for each fixed $D$, the first $D$ terms vanish as $L$ grows; the tail can then be made arbitrarily small by choosing $D$ large.
Thus terminal-outcome noise sensitivity is equivalent to the escape of all outcome variance from every fixed Wiener--Hermite order.

This equivalence concerns approximation by the closed span of Hermite chaoses up to a fixed degree.
It does not rule out arbitrary nonlinear representations, nor is it a computational-complexity statement: an unrestricted representation could simply encode the terminal indicator itself.

\subsection{Macroscopic corollary of Gaussian influence theory}
For a monotone Gaussian event $\mathcal A_L=\{\bm Z:F_L(\bm Z)=1\}$, let $I_i^G(\mathcal A_L)$ denote the geometric influence of coordinate $i$ in the sense of Keller, Mossel, and Sen \cite{SKellerMosselSen2012,SKellerMosselSen2014}.
Assume that the shift coefficient $R>0$ is independent of $L$ and that
\begin{equation}
	F_L(\bm Z+t\bm 1;H)=F_L(\bm Z;H+Rt).
	\label{eq:shiftS}
\end{equation}
Proposition 1.6 of Ref.~\cite{SKellerMosselSen2012}, the Gaussian Russo formula for location families, then gives
\begin{equation}
	\sum_iI_i^G=Rs_L,
	\qquad
	s_L=
	\left.
	\partial_H\mathbb P[F_L(\bm Z;H)=1]
	\right|_{H=H_L^\star}.
\end{equation}

If the event is invariant under a permutation group acting transitively on the coordinates, as in a homogeneous periodic system, then $I_i^G=Rs_L/N$, and hence
\begin{equation}
	\sum_i(I_i^G)^2
	=
	\frac{R^2s_L^2}{L^d}.
	\label{eq:squaredS}
\end{equation}
The microscopic influences have thus been eliminated in favor of the macroscopic response $s_L$.
Set
\begin{equation}
	\mathcal I_L^{(2)}=\sum_i(I_i^G)^2=\frac{R^2s_L^2}{L^d}.
\end{equation}

Care is needed when translating the noise parameter of Ref.~\cite{SKellerMosselSen2014}. 
There the correlated copy is written as
\begin{equation}
	\bm W^\rho=\sqrt{1-\rho^2}\,\bm W+\rho\bm W',
\end{equation}
whereas the Letter uses $\bm Z^{(c)}=c\bm Z+\sqrt{1-c^2}\,\bm\Xi$. Thus $\rho^2=1-c^2$.
Their quantitative Gaussian BKS theorem gives, for universal positive constants $C_1,C_2$,
\begin{equation}
	\operatorname{Cov}[F_L(\bm Z),F_L(\bm Z^{(c)})]
	\le C_1 [\mathcal I_L^{(2)}]^{C_2(1-c^2)}.
	\label{eq:forwardBKS}
\end{equation}
Because the terminal event is balanced, $\mathbb E F_L=1/2$ and hence $\operatorname{Var}(F_L)=1/4$. Normalizing the covariance gives
\begin{equation}
	\mathcal S_L(c)\le 4C_1 [\mathcal I_L^{(2)}]^{C_2(1-c^2)}.
	\label{eq:forwardBKSnorm}
\end{equation}

The reverse (inverse-BKS) lower bound of Ref.~\cite{SKellerMosselSen2014} requires monotonicity. For a monotone Gaussian set it states, in the same parameterization,
\begin{equation}
	\operatorname{Cov}[F_L(\bm Z),F_L(\bm Z^{(c)})]
	\ge c^2\mathcal I_L^{(2)},
	\label{eq:inverseBKS}
\end{equation}
and therefore
\begin{equation}
	\mathcal S_L(c)\ge4c^2\mathcal I_L^{(2)}.
	\label{eq:inverseBKSnorm}
\end{equation}

Combining the two established KMS bounds with the Russo formula and transitivity yields the exact asymptotic corollary
\begin{equation}
	\boxed{
		\frac{s_L}{L^{d/2}}\to0
		\Longleftrightarrow
		\mathcal I_L^{(2)}\to0
		\Longleftrightarrow
		\mathcal S_L(c)\to0
		\quad\text{for every fixed }c\in(0,1).
	}
	\label{eq:iffS}
\end{equation}
Indeed, vanishing of $\mathcal S_L(c)$ for any single fixed $c\in(0,1)$ already forces $\mathcal I_L^{(2)}\to0$ through Eq.~\eqref{eq:inverseBKSnorm}; Eq.~\eqref{eq:forwardBKSnorm} then gives noise sensitivity at every fixed $c\in(0,1)$.
This is not a new probability inequality.
The contribution of the Letter is to identify $s_L$ as a directly measurable physical proxy for $\mathcal I_L^{(2)}$ in terminal-outcome problems.
The equivalence uses independent standard Gaussian inputs, a balanced binary event, coordinatewise monotonicity, uniform-shift equivalence, and coordinate transitivity.
If the shift coefficient depends on size, the corresponding condition is $R_{\mathrm{sh},L}s_L/L^{d/2}\to0$ rather than $s_L/L^{d/2}\to0$.

If $s_L=O(L^{y_F})$ with $y_F<d/2$, the forward quantitative bound can be written schematically as
\begin{equation}
	\mathcal S_L(c)\le
	C(c)L^{-a(c)(d-2y_F)}
\end{equation}
with $a(c)>0$. 
Combining this with
\begin{equation}
	\mathcal S_L(c)\ge(1-\varepsilon)c^{D_\varepsilon(L)}
\end{equation}
gives
\begin{equation}
	D_\varepsilon(L)
	\ge
	\frac{a(c)(d-2y_F)}{|\ln c|}
	\ln L-O_{c,\varepsilon}(1),
\end{equation}
and hence $D_\varepsilon(L)=\Omega(\ln L)$ for every fixed $\varepsilon\in(0,1)$.

\section{RFIM dynamics and response estimates}

\subsection{RFIM dynamics and event-driven implementation}
We simulate
\begin{equation}
	\mathcal H=
	-J\sum_{\langle ij\rangle}S_iS_j
	-\sum_i(H+h_i)S_i,
	\qquad
	h_i=RZ_i,
\end{equation}
on a periodic cubic lattice with $J=1$.
The system starts from $S_i=-1$ for all $i$.

For a down spin, define its current instability threshold
\begin{equation}
	H_i^{\mathrm{th}}
	=
	-h_i-J\sum_{j\in\partial i}S_j,
\end{equation}
the value of the external field at which this spin would flip if $H$ were increased. 
Each down spin has a threshold field determined by its random field and local environment; for fixed $h_i$, each additional up neighbor lowers the threshold by $2J$.
The algorithm advances $H$ directly to the smallest threshold among all down spins. 
The triggering spin flips, and $H$ is held fixed while the avalanche relaxes. 

The production implementation maintains a min-priority queue containing the current threshold of each down spin.
When a neighboring spin flips, the affected threshold is reduced by $2J$ and reinserted with an incremented version counter; stale queue entries are discarded when they are removed from the queue.
This avoids repeated global threshold reconstruction, and no finite field increment $\Delta H$ is introduced.

For each realization we record
\begin{equation}
	H_s=
	\inf\{H:M(H)>M_\star\},
\end{equation}
where $M(H)$ is the magnetization at field $H$ after full avalanche relaxation. Unless stated otherwise, $M_\star=0.9$.
Here $M(H)$ is measured only after all avalanches triggered at field $H$ have fully relaxed.
A separate calibration ensemble determines the size-dependent median $H_L^\star$, ensuring that the same disorder pairs are not used both to set and to test the terminal-event boundary.

\subsection{Production sample sizes and exponent estimates}
The production data at disorder strength $R=2.16$, near the disorder-induced critical region, and $M_\star=0.9$ are summarized in Table~\ref{tab:single}.

\begin{table}[h]
	\caption{Single-world switching-field statistics. $s_L^{\rm KDE}$ is the kernel-density estimate of the switching-field density evaluated at the balanced threshold.}
	\label{tab:single}
	\begin{ruledtabular}
		\begin{tabular}{r r r r r}
			$L$ & samples & median $H_s$ & $s_L^{\rm KDE}$ & $s_L^{\rm KDE}/L^{3/2}$ \\
			\hline
			8   & 20000 & 1.70401 & 2.457   & 0.1086 \\
			12  & 20000 & 1.59411 & 4.252   & 0.1023 \\
			16  & 16000 & 1.54417 & 6.280   & 0.0981 \\
			24  & 12000 & 1.49989 & 11.061  & 0.0941 \\
			32  & 10000 & 1.47917 & 16.101  & 0.0889 \\
			48  & 8000  & 1.46128 & 26.867  & 0.0808 \\
			64  & 6000  & 1.45315 & 40.237  & 0.0786 \\
			80  & 3500  & 1.44878 & 54.285  & 0.0759 \\
			96  & 2500  & 1.44596 & 70.678  & 0.0751 \\
			128 & 1000  & 1.44294 & 99.927  & 0.0690 \\
			160 & 600   & 1.44107 & 126.075 & 0.0623 \\
			192 & 450   & 1.43984 & 169.424 & 0.0637 \\
		\end{tabular}
	\end{ruledtabular}
\end{table}

To verify that the exponent estimate reported in the Letter is not specific to the KDE procedure, we estimate the scaling of the central switching-field distribution in three ways.
The results for $L\ge16$ are summarized in Table~\ref{tab:exp}.
\begin{table}[h]
	\caption{Exponent estimates from sample-level bootstrap resampling.}
	\label{tab:exp}
	\begin{ruledtabular}
		\begin{tabular}{l c c}
			Estimator & $y_F$ & bootstrap $95\%$ CI \\
			\hline
			KDE at median & 1.316 & [1.289,1.348] \\
			inverse IQR & 1.341 & [1.318,1.370] \\
			central 40\% & 1.343 & [1.312,1.372] \\
		\end{tabular}
	\end{ruledtabular}
\end{table}

Repeating the point-estimate fits with different minimum included sizes reveals no systematic drift toward $3/2$ over the available large-size range, supporting robustness with respect to the fitted size range.

\section{RFIM robustness and paired-world simulations}

\subsection{Robustness to the macroscopic switching threshold}	
\begin{figure}[h]
	\includegraphics[width=0.62\textwidth]{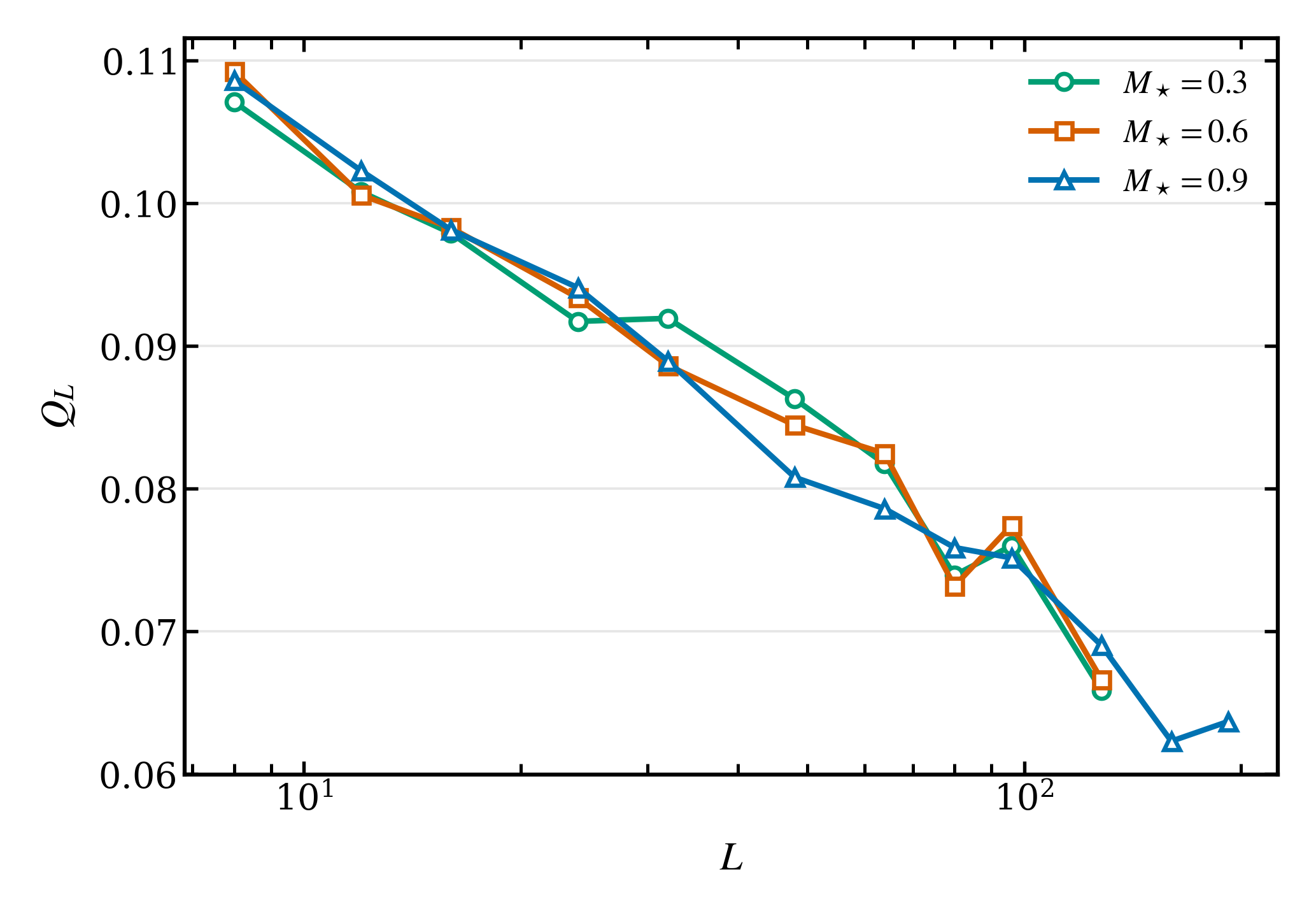}
	\caption{Normalized terminal-event window for three definitions of the macroscopic switching threshold. For all three thresholds, $Q_L=s_L/L^{3/2}$ decreases overall with $L$, showing that the observed sub-$3/2$ scaling is not tied to the choice $M_\star=0.9$.}
	\label{fig:Sthreshold}
\end{figure}

To check that the sub-$3/2$ scaling is not an artifact of the switching threshold, the principal finite-size analysis was repeated for magnetization thresholds $M_\star=0.3$, $0.6$, and $0.9$.
This is illustrated in Fig.~\ref{fig:Sthreshold}, which shows $Q_L$ for the three threshold values; all three curves decrease overall with $L$ and lie close to one another.
Over $L\ge16$, the point estimates from the inverse-IQR and KDE methods lie between approximately $1.32$ and $1.35$ for all three thresholds, and all corresponding bootstrap intervals lie below $3/2$.
Thus the observed sub-$3/2$ scaling is not tied to the choice $M_\star=0.9$.

\subsection{Paired-world simulations and noise strength}
To directly probe the noise sensitivity predicted by the criterion, we measure paired-world correlations at $c=0.99$, corresponding to a weak but nonzero microscopic perturbation.
The number of independent paired disorder realizations and measured correlations are listed in Table~\ref{tab:pair}.
\begin{table}[h]
	\caption{Paired-world terminal-outcome correlations at $c=0.99$.}
	\label{tab:pair}
	\begin{ruledtabular}
		\begin{tabular}{r r c c}
			$L$ & pairs & $\mathcal S_L(0.99)$ & bootstrap $95\%$ CI \\
			\hline
			8   & 10000 & 0.548 & [0.532,0.565] \\
			12  & 10000 & 0.503 & [0.486,0.520] \\
			16  & 8000  & 0.467 & [0.448,0.487] \\
			24  & 7000  & 0.385 & [0.364,0.407] \\
			32  & 6000  & 0.347 & [0.324,0.370] \\
			48  & 5000  & 0.295 & [0.268,0.321] \\
			64  & 4000  & 0.265 & [0.235,0.294] \\
			80  & 3000  & 0.230 & [0.195,0.264] \\
			96  & 2400  & 0.226 & [0.186,0.265] \\
			128 & 1600  & 0.257 & [0.209,0.303] \\
			160 & 1000  & 0.186 & [0.127,0.248] \\
			192 & 700   & 0.203 & [0.129,0.278] \\
		\end{tabular}
	\end{ruledtabular}
\end{table}
The correlation decreases overall with $L$, providing finite-size evidence consistent with noise sensitivity even for a disorder correlation as large as $c=0.99$.

\begin{figure}[h]
	\includegraphics[width=0.62\textwidth]{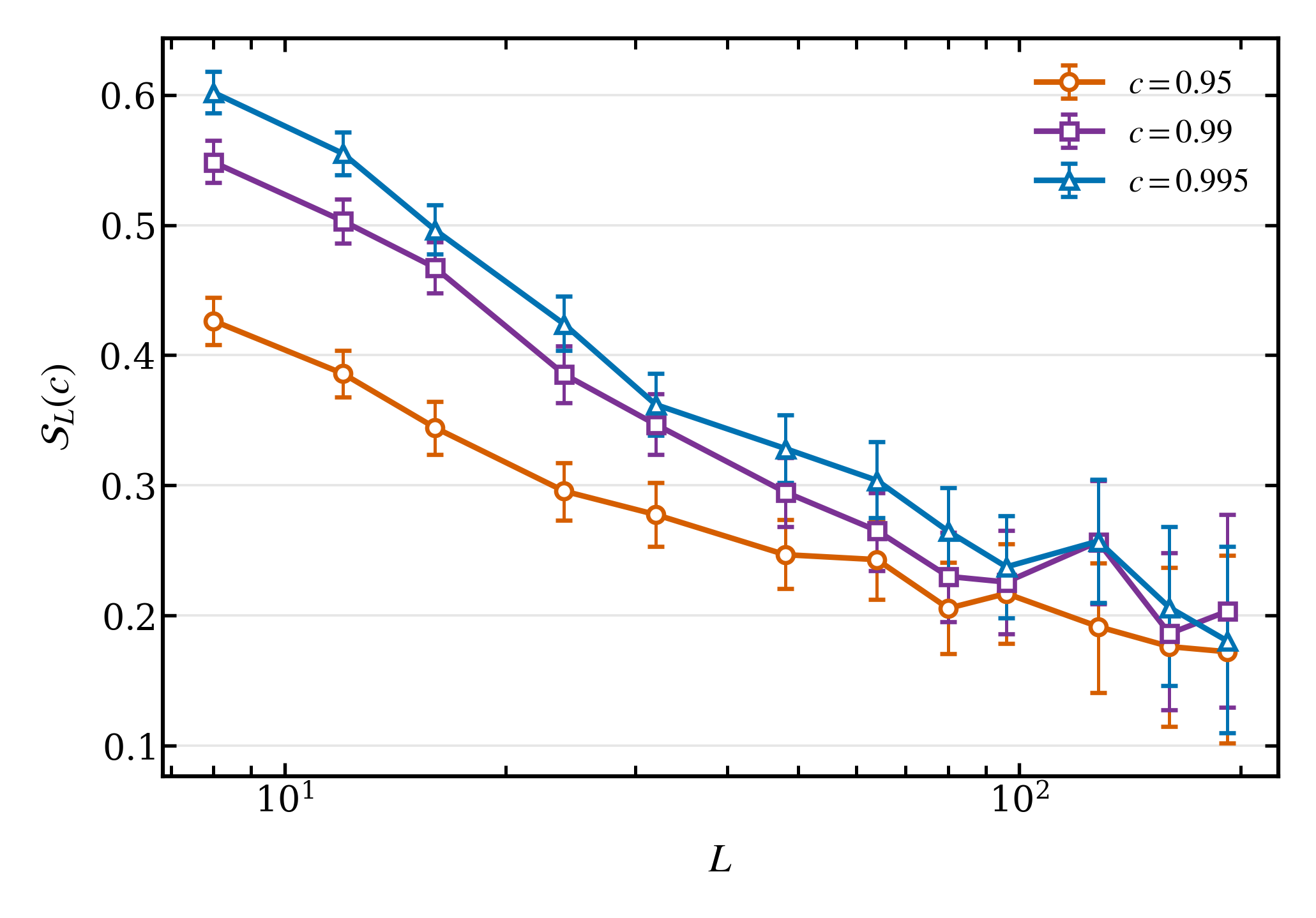}
	\caption{Paired-world terminal-outcome correlations $\mathcal S_L(c)$ for several microscopic disorder correlations $c$. For each fixed $c$, the correlation decreases overall with system size, and the ordering $\mathcal S_L(0.995)>\mathcal S_L(0.99)>\mathcal S_L(0.95)$ holds within statistical uncertainty.}
	\label{fig:Snoise}
\end{figure}

Additional calculations at $c=0.95$ and $0.995$ show the expected ordering with noise strength and the same overall decrease with system size, as shown in Fig.~\ref{fig:Snoise}.

It should be emphasized that these finite-size curves are used as an operational consistency test, not as proof that $\mathcal S_L(c)\to0$ in the thermodynamic limit.
In particular, for $c$ close to unity the quantitative Gaussian BKS bound permits a slow crossover.

\subsection{Confirmatory finite-noise crossover}
We used a staged, prospectively frozen protocol---i.e., the analysis choices for each stage were fixed before the data for that stage were generated---to measure the noise scale at which the balanced terminal-outcome correlation crosses one half.
For each size
\begin{equation}
	L\in\{24,32,48,64\}
\end{equation}
and disorder
\begin{equation}
	R\in\{2.145,2.150,2.155,2.160,2.165,2.170,2.175\},
\end{equation}
an independent calibration sample fixed the balanced terminal-event threshold.
A separate 500-row scan was used to bracket the crossover; subsequently, 1800 new shared-disorder rows were generated on a fixed grid of correlated worlds.
All noise levels within a row share the same base disorder and Gaussian innovation.
Uncertainty is therefore obtained by resampling complete rows.
We project the measured correlation monotonically in $\eta=1-c^2$---the variance of the independent Gaussian innovation---and interpolate in $\log\eta$ to define
\begin{equation}
	\mathcal S_L(\sqrt{1-\eta_{1/2}})=\frac12.
\end{equation}

Every final condition yields a finite half point in all 1000 bootstrap draws, with exact agreement between the $c=1$ and base switching fields, and at least two projected grid points on each side of one half. 
The final family retains $26$ v1.2 conditions and includes independently generated v1.3 replacement samples for $(L,R)=(24,2.155)$ and $(48,2.175)$; the corresponding failed v1.2 samples were excluded from inference.
The replacement grids were frozen before the replacement samples were generated, and every condition in the final family is bracketed by its prespecified nonzero grid.
In the final 28-condition family, a two-sided exact-binomial base-balance test (Holm family level $0.05$) gives no rejection. Two-sided Fisher tests for independence between base and $c=0$ fates give no Holm rejection at family level $0.025$; their global Fisher combination gives $p_{\rm global}=0.7548$. 
No calibration repair was triggered.

The final point estimates are listed in Table~\ref{tab:halfnoise}; complete-row bootstrap intervals are shown in Fig.~2 of the Letter.
For fixed $R$, $\eta_{1/2}$ decreases systematically with $L$, indicating that larger systems require less noise to destroy half of the correlation.
\begin{table}[h]
	\caption{Confirmatory half-correlation noise scales $\eta_{1/2}$ as functions of size and disorder strength.}
	\label{tab:halfnoise}
	\begin{ruledtabular}
		\begin{tabular}{r c c c c c c c}
			$L$ & $2.145$ & $2.150$ & $2.155$ & $2.160$ & $2.165$ & $2.170$ & $2.175$ \\
			\hline
			24 & $3.22\!\times\!10^{-3}$ & $3.61\!\times\!10^{-3}$ & $3.61\!\times\!10^{-3}$ & $2.41\!\times\!10^{-3}$ & $2.24\!\times\!10^{-3}$ & $1.88\!\times\!10^{-3}$ & $3.90\!\times\!10^{-3}$ \\
			32 & $1.50\!\times\!10^{-3}$ & $1.22\!\times\!10^{-3}$ & $1.04\!\times\!10^{-3}$ & $1.55\!\times\!10^{-3}$ & $1.74\!\times\!10^{-3}$ & $9.17\!\times\!10^{-4}$ & $1.29\!\times\!10^{-3}$ \\
			48 & $4.31\!\times\!10^{-4}$ & $5.07\!\times\!10^{-4}$ & $5.14\!\times\!10^{-4}$ & $6.42\!\times\!10^{-4}$ & $7.44\!\times\!10^{-4}$ & $5.89\!\times\!10^{-4}$ & $4.63\!\times\!10^{-4}$ \\
			64 & $1.84\!\times\!10^{-4}$ & $1.42\!\times\!10^{-4}$ & $2.30\!\times\!10^{-4}$ & $1.89\!\times\!10^{-4}$ & $1.97\!\times\!10^{-4}$ & $2.45\!\times\!10^{-4}$ & $1.35\!\times\!10^{-4}$ \\
		\end{tabular}
	\end{ruledtabular}
\end{table}

To test for universality, we also examined whether the half-correlation data collapse onto a single scaling surface. 
The prospectively specified form was
\begin{align}
	\log\eta_{1/2}
	&=\alpha-\zeta\log(L/48)+b_1y+b_2y^2,
	\\
	y
	&=\frac{R-R_c}{0.01}\left(\frac{L}{48}\right)^{1/\nu},
\end{align}
with 120 Latin-hypercube multistarts and bounds
\begin{equation}
	R_c\in[2.13,2.19],
	\qquad
	\nu\in[0.5,3].
\end{equation}
The optimum placed $R_c=2.190$ on its upper bound. Its AICc is $37.25$, whereas a two-parameter no-$R$ baseline has AICc $29.44$, giving
\begin{equation}
	\Delta{\rm AICc}
	={\rm AICc}_{\rm no\text{-}R}-{\rm AICc}_{\rm two\text{-}variable}
	=-7.80.
\end{equation}
The frozen acceptance threshold was $\Delta{\rm AICc}\ge6$. 
Moreover, six of seven leave-one-disorder fits reach a parameter boundary, and the largest shift in $R_c$ is $0.060$, exceeding the prespecified maximum of $0.008$. 
We therefore do not report $R_c$, $\nu$, or $\zeta$ as identified crossover parameters. 
Figure~2 of the Letter intentionally presents the raw finite-size narrowing without a universal collapse.

\subsection{Off-critical control}
As an off-critical control, we repeat the switching-width analysis at $R=3.0$.
Over $L=16$--$128$, the inverse interquartile switching width gives $y_F=1.471$ with bootstrap $95\%$ interval $[1.441,1.498]$, close to the square-root-volume value $3/2$. Correspondingly, the quantity
\begin{equation}
	\frac{[\mathrm{IQR}(H_s)]^{-1}}{L^{3/2}}
\end{equation}
is approximately size independent over the largest sizes studied, in contrast to the $R=2.16$ sequence near the disorder-induced critical region.

This control is consistent with the sub-$L^{3/2}$ terminal-event-window scaling at $R=2.16$ being associated with the critical region rather than being a generic effect of independent Gaussian disorder.

\section{Evolutionary-game implementation}
The criterion extends beyond magnetism to a monotone spatial Stag-Hunt adoption process.
For a lattice agent with $k_i$ cooperative neighbors, the cooperation payoff advantage is
\begin{equation}
	\Delta u_i=H+\sigma Z_i+k_i(r-t)+(z-k_i)(s-p)
	+\kappa\frac{\binom{k_i}{2}}{\binom z2},
	\label{eq:stagS}
\end{equation}
where $r>t\ge p>s$, $\sigma>0$, and $\kappa\ge0$ \cite{SPachecoEtAl2009}.
The first four terms are analogous to the RFIM local field; the $\kappa$-term introduces a genuine multi-agent coordination bonus.
Starting from prescribed committed seeds, $H$ is increased quasistatically, and every strict defector-to-cooperator best response is retained.
Strategic complementarity ensures that the resulting monotone closure is independent of the update order \cite{SWatts2002}.
Strategic complementarity follows from
\begin{equation}
	\Delta u_i(k+1)-\Delta u_i(k)
	=r-t-s+p+\frac{2\kappa k}{z(z-1)}>0.
	\label{eq:complementS}
\end{equation}

Nontrivial prescribed seeds generally break coordinate transitivity, so the geometric influences need not be equal.
In the transitive case, the equal-influence property was essential for reducing the squared influences $\sum_i(I_i^G)^2$ to $R^2s_L^2/L^d$. Without it, we must retain the full distribution of influences. 
This is naturally captured by the effective number of influential coordinates, which reduces to the volume when all influences are equal and to a small number when the response is concentrated.

For a balanced terminal cooperation-level event, we introduce the effective number of influential preference coordinates,
\begin{equation}
	N_{\mathrm{eff},L,q}
	=\frac{\left(\sum_i I_{i,L,q}^G\right)^2}{\sum_i (I_{i,L,q}^G)^2}.
\end{equation}
By Cauchy--Schwarz, $1\le N_{\mathrm{eff},L,q}\le L^d$, with equality to $L^d$ when all influences are equal.
By the same Russo identity, the sum of influences is still $\sigma s_{L,q}$, while the sum of their squares becomes $\sigma^2 s_{L,q}^2/N_{\mathrm{eff},L,q}$. 
Substituting this into the KMS bounds yields the natural generalization of the criterion, with $L^{d/2}$ replaced by $\sqrt{N_{\mathrm{eff},L,q}}$:
\begin{equation}
	\mathcal S_{L,q}(c)\to0
	\quad\text{for every fixed }c\in(0,1)
	\Longleftrightarrow
	\quad
	\frac{\sigma s_{L,q}}{\sqrt{N_{\mathrm{eff},L,q}}}\to0.
	\label{eq:gamecriterionS}
\end{equation}
Thus coordinate transitivity closes the criterion in terms of the global slope alone; with arbitrary seeds, one must also determine how concentrated the site-resolved response is.
For $\kappa=0$, the update rule maps sample by sample onto a ferromagnetic zero-temperature RFIM, with the committed seeds represented as pinned up spins \cite{SZimmaroEtAl2024}; $\kappa>0$ introduces a genuine higher-order assurance interaction.

The Stag-Hunt process is monotone and therefore removes path dependence by construction. 
To complement this, we study a distinct bidirectional asynchronous best-response game with an exact potential, in which different random strict-improvement schedules can reach different Nash equilibria from the same initial state \cite{SBlume1995,SMondererShapley1996}.
For $L=8$ to $48$, the fraction of binary-outcome variance attributable to the initial state remains between $0.554$ and $0.572$, while approximately $67\%$ of the initial states at the two largest sizes produce both outcomes under different schedules, indicating that the initial state carries substantial outcome information even in the presence of schedule-dependent equilibrium multiplicity.
A mixed Hoeffding--Hermite analysis gives curvewise-bootstrap lower bounds of at least $0.616$ and $0.708$ for the outcome variance contained through orders $10$ and $20$, respectively, at $L=32$ and $48$.
These finite-size results show that schedule-dependent equilibria can coexist with a substantial finite-order predictive component; they are not intended as thermodynamic-limit claims.

\subsection{Prescribed initial seeds and final-macrostate information}
We now generalize beyond the translation-invariant initial state. 
Let $\bm a_L^{(0)}\in\{0,1\}^{L^d}$ be an arbitrary prescribed seed configuration; its cooperators are retained as committed seeds. 
At fixed external field $(H,\bm Z)$, repeatedly adopt every eligible defector until no further adoption is possible. Equation~\eqref{eq:complementS} makes this a monotone closure, so every fair update order reaches the same least fixed point $\bm a_L^\infty(\bm Z;\bm a_L^{(0)},H)$ above the seed state. 
Define
\begin{equation}
	X_L^\infty(\bm Z;\bm a_L^{(0)},H)
	=\frac{1}{L^d}\sum_i a_{L,i}^\infty
\end{equation}
and, for a fixed cooperation level $q$ above the seed density,
\begin{equation}
	F_{L,q}(\bm Z;\bm a_L^{(0)},H)
	=\bm 1\!\left\{X_L^\infty(\bm Z;\bm a_L^{(0)},H)\ge q\right\}.
	\label{eq:gameLevelS}
\end{equation}
Increasing any $Z_i$ can only enlarge the closure, so every level-set event is coordinatewise monotone. 
When the event is nondegenerate, let $H_{L,q}^\star$ be its balancing field, let $s_{L,q}$ be the slope of the switching probability with respect to $H$ at that field, and let $I_{i,L,q}^G$ be the corresponding Gaussian geometric influences.

The loss of translation symmetry makes the influences unequal, but they remain operationally accessible through local incentive responses. 
Add a site-specific incentive $\delta_i$ to the payoff difference of agent $i$ and define
\begin{equation}
	g_{i,L,q}
	=\left.
	\frac{\partial}{\partial\delta_i}
	\mathbb P(F_{L,q}=1)
	\right|_{\bm\delta=0,\,H=H_{L,q}^\star}.
\end{equation}
This quantity is the local response of the terminal probability to a change in agent $i$'s personal incentive.
Shifting $Z_i$ by $t$ is equivalent to setting $\delta_i=\sigma t$; hence the one-coordinate Gaussian Russo identity and the global response give
\begin{equation}
	I_{i,L,q}^G=\sigma g_{i,L,q},
	\qquad
	s_{L,q}=\sum_i g_{i,L,q},
	\qquad
	\sum_i I_{i,L,q}^G=\sigma s_{L,q}.
	\label{eq:localResponseS}
\end{equation}
Define the effective number of influential agents by
\begin{equation}
	N_{\mathrm{eff},L,q}
	=\frac{\left(\sum_i g_{i,L,q}\right)^2}
	{\sum_i g_{i,L,q}^2}
	=\frac{\left(\sum_i I_{i,L,q}^G\right)^2}
	{\sum_i(I_{i,L,q}^G)^2}.
	\label{eq:neffS}
\end{equation}
Nonnegativity and Cauchy--Schwarz give $1\le N_{\mathrm{eff},L,q}\le L^d$; equality on the right holds when all agents
have equal influence. 
Combining Eqs.~\eqref{eq:localResponseS} and \eqref{eq:neffS} eliminates the individual influences in favor of the effective number of coordinates:
\begin{equation}
	A_{L,q}:=\sum_i(I_{i,L,q}^G)^2
	=\frac{\sigma^2s_{L,q}^2}{N_{\mathrm{eff},L,q}}.
	\label{eq:generalAS}
\end{equation}
Applying the normalized forward and inverse KMS bounds level by level yields
\begin{equation}
	\boxed{
		\mathcal S_{L,q}(c)\to0\ \text{for every fixed }c\in(0,1)
		\quad\Longleftrightarrow\quad
		\frac{\sigma s_{L,q}}{\sqrt{N_{\mathrm{eff},L,q}}}\to0.
	}
	\label{eq:generalGameCriterionS}
\end{equation}
Thus the departure from transitivity introduced by localized seeds enters the criterion through the concentration of the site-resolved response. 
For a translation-invariant initial state, $N_{\mathrm{eff},L,q}=L^d$, and Eq.~\eqref{eq:generalGameCriterionS} becomes
$\sigma s_{L,q}/L^{d/2}\to0$, recovering the transitive criterion with $R$ replaced by $\sigma$.

The level-set family contains the complete distribution of the final macroscopic state. 
At any fixed $H$ and prescribed seed state, write $T_{L,H}(q)=\mathbb P(X_L^\infty\ge q)$. With $N=L^d$,
\begin{align}
	\mathbb P\!\left(X_L^\infty=\frac{k}{N}\right)
	&=T_{L,H}\!\left(\frac{k}{N}\right)
	-T_{L,H}\!\left(\frac{k+1}{N}\right),
	\label{eq:distributionRecoveryS}\\
	\mathbb E[X_L^\infty]
	&=\frac1N\sum_{k=1}^{N}T_{L,H}\!\left(\frac{k}{N}\right),
\end{align}
where $T_{L,H}((N+1)/N)=0$. 
The balanced slopes diagnose the noise sensitivity of these level sets; the unbalanced tails reconstruct the final-state
distribution. 
For a fully specified deterministic pair $(\bm a_L^{(0)},\bm Z)$ the distribution is degenerate. 
The criterion describes robustness under a specified ensemble of preference perturbations and the failure of fixed Gaussian-chaos order; it is not an algorithm for obtaining an individual final state without performing the monotone closure.

For $\kappa=0$, set $S_i=2a_i-1$ and $b=r-t-s+p$. Equation~\eqref{eq:stagS} becomes
\begin{equation}
	\Delta u_i
	=H+\sigma Z_i+\frac b2\sum_{j\in\partial i}S_j
	+\frac z2(r-t+s-p),
	\label{eq:rfimMappingS}
\end{equation}
where the last term is the constant contribution from the baseline payoff values.
This is the zero-temperature RFIM local-field condition with
\begin{equation}
	J=\frac b2,
	\qquad
	h_i=\sigma Z_i,
	\qquad
	H_{\rm RFIM}=H+\frac z2(r-t+s-p),
\end{equation}
consistent with known mappings between asymmetric coordination games and random-field spin systems \cite{SZimmaroEtAl2024}. 
We use this equality as a sample-by-sample validation of the event-driven code. 
For the numerical payoff differences $r-t=0.4$ and $p-s=0.6$ on the cubic lattice with $z=6$, one has $J=1/2$ and $H_{\rm RFIM}=H-0.6$; the two implementations agree to floating-point precision for identical disorder samples.
At $\kappa>0$, the term quadratic in $k_i$ generates multi-agent interactions and is not an ordinary pairwise RFIM, while monotonicity and uniform-shift equivalence remain intact; coordinate transitivity still depends on the seed pattern.

The event-driven implementation stores the incentive threshold for each defector:
\begin{equation}
	H_i(k)=-\sigma Z_i-k(r-t)-(z-k)(s-p)
	-\kappa\frac{\binom{k}{2}}{\binom z2}.
\end{equation}
The smallest current threshold triggers the next avalanche. 
Each cooperative neighbor lowers the thresholds of affected defectors via Eq.~\eqref{eq:complementS}; all newly eligible vertices are processed at the same $H$. 
This continues until full cooperation. 
The switching field and paired-world switching fields are thus obtained without discretizing $H$.

\subsection{Direct measurement of a nontransitive response}
We evaluate the generalized criterion on a periodic cubic lattice with $z=6$, $r-t=0.4$, $p-s=0.6$, $\sigma=1.45$, $\kappa=2$, and terminal level $q=0.25$.
All $L^2$ sites in one coordinate plane are committed cooperators, leaving
\begin{equation}
	N_{\mathrm{free}}=L^3-L^2
\end{equation}
independent Gaussian preference coordinates.
This seed breaks translation symmetry in the direction normal to the plane and therefore produces a genuinely nontransitive influence profile.
The balancing field is independently calibrated at each size; the pooled terminal-event frequency remains close to one half, as reported in Table~\ref{tab:nontransitiveResponseS}.

The local responses can be estimated without applying a separate perturbation to every coordinate.
At fixed $H$, Gaussian integration by parts gives the score identity
\begin{equation}
	g_{i,L,q}
	=\frac{1}{\sigma}
	\mathbb E\!\left[
	\bigl(F_{L,q}-\mathbb E F_{L,q}\bigr)Z_i
	\right]
	\label{eq:scoreResponseS}
\end{equation}
for every free coordinate $i$.
We use two statistically independent production blocks at each size.
Each block contains $20\,000$ disorder realizations in $80$ chunks for $L=8,12,16$, $16\,000$ realizations in $64$ chunks for $L=20$, and $12\,000$ realizations in $48$ chunks for $L=24$.
If $\widehat g_i^{(0)}$ and $\widehat g_i^{(1)}$ are the two score estimates and $\widehat s_b=\sum_i\widehat g_i^{(b)}$, the cross-block estimators
\begin{align}
	\widehat Q
	&=\sum_{i\in\mathrm{free}}
	\widehat g_i^{(0)}\widehat g_i^{(1)},
	\label{eq:crossQS}\\
	\widehat N_{\mathrm{eff}}
	&=\frac{\widehat s_0\widehat s_1}{\widehat Q},
	\qquad
	\widehat A=\sigma^2\widehat Q
	\label{eq:crossNeffS}
\end{align}
remove the positive noise-square bias that would arise from squaring a single-block estimate.
At event probability $p_L$, the normalized first-order Hermite weight is
\begin{equation}
	W_{L,1}=\frac{A_{L,q}}{p_L(1-p_L)}.
	\label{eq:firstChaosScoreS}
\end{equation}
Chunk-bootstrap resampling is performed separately within the two blocks before the cross-products are formed.

The response vector also has the exact orthogonal decomposition
\begin{equation}
	\sum_{i\in\mathrm{free}}g_i^2
	=\frac{s_{L,q}^2}{N_{\mathrm{free}}}
	+\left\lVert
	\bm g-\frac{s_{L,q}}{N_{\mathrm{free}}}\bm 1
	\right\rVert_2^2.
	\label{eq:responseDecompositionS}
\end{equation}
The second term isolates the part of the squared response that cannot be represented by a uniform influence profile.
For committed seeds, Cauchy--Schwarz therefore sharpens the general upper bound to $N_{\mathrm{eff}}\le N_{\mathrm{free}}$.

\begin{table}[t]
	\caption{Direct score estimates for the plane-seeded monotone game. Brackets are chunk-bootstrap $95\%$ intervals. The intervals for $W_{L,1}$ normalize the corresponding $A$ intervals by the measured Bernoulli variance. Here $Q_\perp$ is the second term in Eq.~\eqref{eq:responseDecompositionS}.}
	\label{tab:nontransitiveResponseS}
	\begin{ruledtabular}
		\begin{tabular}{c c c c c c}
			$L$ & $H_{L,q}^\star$ & $p_L$ & $N_{\rm eff}/N_{\rm free}$ & $W_{L,1}$ & $Q_\perp$ \\
			\hline
			8  & $0.5814$ & $0.4919$ & $0.582\,[0.562,0.601]$ & $0.289\,[0.280,0.300]$ & $0.0144\,[0.0135,0.0153]$ \\
			12 & $0.7252$ & $0.5010$ & $0.568\,[0.544,0.593]$ & $0.296\,[0.284,0.309]$ & $0.0152\,[0.0140,0.0165]$ \\
			16 & $0.7928$ & $0.4907$ & $0.602\,[0.574,0.634]$ & $0.283\,[0.269,0.296]$ & $0.0134\,[0.0119,0.0149]$ \\
			20 & $0.8349$ & $0.5072$ & $0.596\,[0.551,0.644]$ & $0.260\,[0.241,0.280]$ & $0.0125\,[0.0103,0.0148]$ \\
			24 & $0.8619$ & $0.5054$ & $0.649\,[0.577,0.739]$ & $0.262\,[0.229,0.292]$ & $0.0109\,[0.0072,0.0145]$
		\end{tabular}
	\end{ruledtabular}
\end{table}

The effective influence fraction remains between $0.568$ and $0.649$ across the five sizes.
A descriptive power-law fit over the five available sizes gives
\begin{equation}
	N_{\mathrm{eff}}/N_{\mathrm{free}}
	\propto L^{0.089},
	\qquad 95\%\ \mathrm{CI}=[0.004,0.182],
	\label{eq:neffFractionFitS}
\end{equation}
while a separate descriptive fit to $Q_\perp$ gives an exponent of $-0.252\,[-0.531,-0.050]$.
The squared-response fraction contained within the first two layers away from the seed plane decreases from $0.977$ at $L=8$ to $0.660$ at $L=24$, providing finite-size evidence that the response broadens into the bulk rather than remaining confined to the immediate vicinity of the seed plane.
Together, the order-one effective influence fraction and substantial $W_{L,1}$ provide a finite-size example in which broken transitivity is directly resolved but a low-order predictive core persists.
Only five sizes are available, so these measurements are not used to assert a thermodynamic limit.
All expected chunks are present at each size, the two blocks contain disjoint random seeds, and the seed masks and calibration files pass the recorded hash and range checks.

\subsection{Path-dependent asynchronous game and initial-state determinacy}
The monotone closure above isolates the hypotheses of the slope criterion, but removes update-order dependence by construction. 
To probe the predictive framework beyond the no-passing class, we therefore also study a bidirectional asynchronous Stag-Hunt process in which the same initial state can reach different terminal equilibria under different schedules.
This second model is not used as evidence for Eq.~\eqref{eq:generalGameCriterionS}; rather, it tests the coexistence of schedule dependence with finite-order predictability.

On a periodic square lattice, let $a_i\in\{0,1\}$ and let $k_i$ be the number of cooperative neighbors. 
The cooperation payoff advantage is
\begin{equation}
	\Delta_i(k_i)
	=H+\sigma Z_i-\alpha
	+(1-\beta+\alpha)\frac{k_i}{4},
	\label{eq:reversibleGameS}
\end{equation}
with $\alpha,\beta\in(0,1)$ and $Z_i\sim\mathcal N(0,1)$. 
Initially, strategies are independent Bernoulli variables with density $\rho_0$. 
At each event, one currently unstable site is chosen uniformly and makes a strict best response. 
The game is an exact potential game, and its strict best-response dynamics ascends the potential.
Indeed, with $b=(1-\beta+\alpha)/4$ and $h_i=H+\sigma Z_i-\alpha$, define
\begin{equation}
	\Phi(\bm a)=\sum_i h_i a_i+b\sum_{\langle ij\rangle}a_i a_j.
	\label{eq:potentialS}
\end{equation}
A flip $0\to1$ changes $\Phi$ by $\Delta_i(k_i)$; a flip $1\to0$ changes it by $-\Delta_i(k_i)$.
Every strict best response therefore increases $\Phi$ by the agent's payoff improvement; because the state space is finite, every run terminates at a pure-strategy Nash equilibrium.
The terminal equilibrium need not be unique because the order of updates matters.

Let
\begin{equation}
	\mathcal X_L=(\bm a^{(0)},\bm Z),\qquad
	Y_L=\bm 1\!\left\{L^{-2}\sum_i a_i^\infty\ge q\right\},
\end{equation}
and denote the random update schedule by $\Omega$. 
The optimal initial-state predictor under squared loss is
\begin{equation}
	g_L(\mathcal X_L)=\mathbb E_\Omega[Y_L\mid\mathcal X_L].
\end{equation}
We quantify how much of the terminal-outcome variance is attributable to the complete initial state by
\begin{equation}
	D_L=
	\frac{\operatorname{Var}_{\mathcal X}[g_L(\mathcal X_L)]}
	{\operatorname{Var}_{\mathcal X,\Omega}(Y_L)}.
	\label{eq:determinacyS}
\end{equation}
Thus $D_L=0$ means that the initial state carries no information about the binary terminal outcome, whereas $D_L=1$ means that the update schedule does not change that outcome.

For $M$ independently drawn initial states and $n_\Omega$ schedules per state, define
\begin{equation}
	B=\operatorname{Var}_m(\bar Y_m),\qquad
	V_W=\frac1M\sum_m\operatorname{Var}_r(Y_{mr}),
\end{equation}
where $B$ is the between-state variance of the schedule-averaged outcomes and $V_W$ is the within-state variance. 
The equal-replicate random-effects estimator is
\begin{equation}
	\widehat{D}_L=
	\frac{B-V_W/n_\Omega}{B+(1-1/n_\Omega)V_W}.
	\label{eq:anovaDeterminacyS}
\end{equation}
Complete initial-state rows, rather than individual schedules, are resampled in the cluster bootstrap.

To resolve the information order of the initial-state predictor, we now measure how its stability decays under perturbation.
Construct a paired initial state $\mathcal X_L^{(c)}$: each Bernoulli coordinate is retained with probability $c$ and otherwise independently redrawn, while
\begin{equation}
	Z_i^{(c)}=cZ_i+\sqrt{1-c^2}\,\xi_i,
	\qquad \xi_i\sim\mathcal N(0,1).
\end{equation}
Define the normalized stability
\begin{equation}
	\mathcal R_L(c)=
	\frac{\operatorname{Cov}\!\left[g_L(\mathcal X_L),
		g_L(\mathcal X_L^{(c)})\right]}
	{\operatorname{Var}[g_L(\mathcal X_L)]}.
	\label{eq:gameStabilityS}
\end{equation}
The product Bernoulli--Gaussian noise operator diagonalizes in the mixed Hoeffding--Hermite basis. 
Writing $w_m(L)$ for the fraction of
$\operatorname{Var}[g_L(\mathcal X_L)]$ carried by total degree $m$ in the centered predictor
$g_L-\mathbb E[g_L]$, we obtain
\begin{equation}
	\mathcal R_L(c)=\sum_{m\ge1}w_m(L)c^m,
	\qquad w_m(L)\ge0,
	\qquad \sum_{m\ge1}w_m(L)=1.
	\label{eq:mixedSpectrumS}
\end{equation}
For
$W_{L,K}=\sum_{m=1}^K w_m(L)$,
every $c\in(0,1)$ therefore gives
\begin{equation}
	\max\!\left\{0,
	\frac{\mathcal R_L(c)-c^{K+1}}{1-c^{K+1}}\right\}
	\le W_{L,K}\le
	\min\!\left\{1,\frac{\mathcal R_L(c)}{c^K}\right\}.
	\label{eq:spectralBoundsS}
\end{equation}

We use $(\alpha,\beta,\sigma,\rho_0,q)=(0.35,0.40,0.35,0.50,0.50)$ and independently calibrate $H_L$ to target a balanced binary terminal outcome.
Every production size uses $M=1000$ initial states and $n_\Omega=20$ schedules.
Paired-world covariances are estimated from ten independent schedules per member.
For each of the 5000 complete-row bootstrap resamples at $L=32,48$, we impose $\mathcal R_L(0)=0$, $\mathcal R_L(1)=1$, and monotonicity before intersecting the bounds obtained from all measured $c$ values. 
Incompatible bootstrap intersections are retained rather than discarded; their frequency is reported as a diagnostic.

Across all six sizes, $D_L$ ranges from $0.554$ to $0.572$, and the bootstrap intervals overlap throughout the simulated range.
At $L=32$ and $48$, $D_L=0.564\,[0.539,0.588]$ and $0.557\,[0.531,0.581]$, respectively.
At the same two sizes, $68.2\%$ and $66.7\%$ of initial states produce more than one binary outcome across their twenty schedules.
Hence the observed intermediate determinacy is not an artifact of path-independent outcomes.

\begin{table}[t]
	\caption{Largest-size path-dependent game results. Spectral entries are lower ends of curvewise bootstrap $95\%$ envelopes; ``valid'' is the fraction of bootstrap curves for which the lower bound does not exceed the upper bound after intersection.}
	\label{tab:gameDeterminacyS}
	\begin{ruledtabular}
		\begin{tabular}{ccccccc}
			$L$ & $H_L$ & $D_L$ & path dep. & $W_{10}^{\rm lo}$ & valid & $W_{20}^{\rm lo}$ \\
			\hline
			32 & $-0.12538$ & $0.564\,[0.539,0.588]$ & $0.682$ & $0.616$ & $1.000$ & $0.708$ \\
			48 & $-0.12506$ & $0.557\,[0.531,0.581]$ & $0.667$ & $0.627$ & $1.000$ & $0.709$
		\end{tabular}
	\end{ruledtabular}
\end{table}

\begin{figure}[t]
	\includegraphics[width=\textwidth]{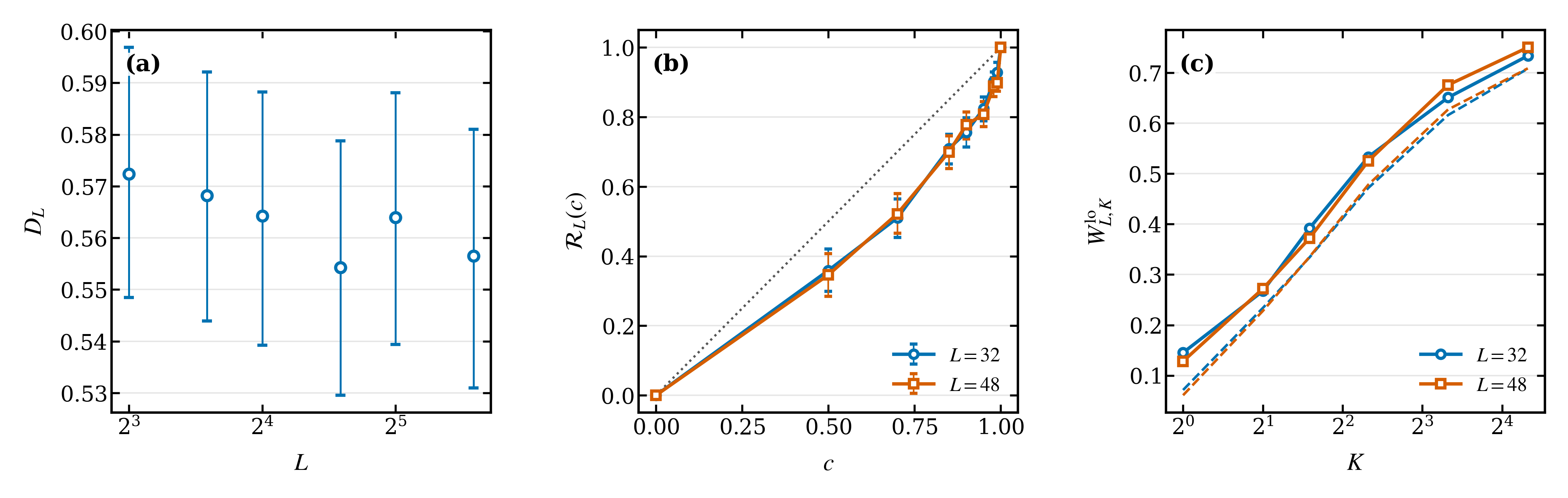}
	\caption{Path-dependent evolutionary-game validation. (a) Initial-state determinacy remains at an intermediate value over the simulated sizes; error bars are cluster-bootstrap $95\%$ intervals. (b) Endpoint-constrained isotonic estimates of the normalized initial-state stability; error bars are pointwise projected-bootstrap $95\%$ intervals. (c) Intersections of the spectral lower bounds over all measured $c$ values. Filled symbols are point bounds and open dashed symbols are the lower ends of the curvewise bootstrap envelopes.}
	\label{fig:gameDeterminacyS}
\end{figure}

The corresponding stability curves and spectral bounds are shown in Fig.~\ref{fig:gameDeterminacyS}. 
For both $K=10$ and $K=20$, the compatibility fraction is $1.000$ at both sizes. 
Thus the largest-size data support the finite-size statements, at the curvewise bootstrap-envelope level,
\begin{equation}
	W_{L,10}\gtrsim0.62,
	\qquad
	W_{L,20}\gtrsim0.71,
	\qquad L\in\{32,48\},
	\label{eq:gameSpectralResultS}
\end{equation}
where the inequalities are understood in the sense of the lower ends of the bootstrap envelopes. These are not thermodynamic lower bounds. 
They show that a substantial, noise-stable finite-order predictive core coexists with persistent schedule dependence over the simulated range.

\section{Bethe-lattice two-world benchmark}
To separate mean-field criticality from the finite-dimensional information criterion used in the Letter, we consider a Bethe lattice benchmark. 
For coordination $z$, let $b=z-1$ and let $P_n$ denote the probability that a cavity site flips up while its parent is held down. 
If exactly $m$ of its $b$ descendants are up, define the conditional flip probability
\begin{equation}
	p_m(H)=
	\mathbb P\!\left[RZ+H+J(2m-z)>0\right]
	=\Phi_{\rm G}\!\left(\frac{H+J(2m-z)}{R}\right),
\end{equation}
where $\Phi_{\rm G}$ is the standard-normal cumulative distribution function; below we write $p_m=p_m(H)$.
The exact recursion is \cite{SDharShuklaSethna1997}
\begin{equation}
	P_{n+1}
	=
	G(P_n)
	=
	\sum_{m=0}^{b}
	\binom bm
	P_n^m(1-P_n)^{b-m}p_m.
	\label{eq:betheG}
\end{equation}
Consider two copies with identical local disorder and the same one-world marginal $P_n$, but with a dilute set of independently distributed disagreements among corresponding descendant states.
If $D_n$ denotes the descendant disagreement probability, expansion to first order in $D_n$ gives
\begin{equation}
	D_{n+1}
	=
	\Lambda(P_n)D_n+O(D_n^2),
\end{equation}
where
\begin{equation}
	\Lambda(P)
	=
	b
	\sum_{m=0}^{b-1}
	\binom{b-1}{m}
	P^m(1-P)^{b-1-m}
	(p_{m+1}-p_m).
\end{equation}
The quantity $\Lambda(P)$ governs the propagation of a microscopic disagreement through the tree.
Differentiating Eq.~\eqref{eq:betheG} yields the exact identity
\begin{equation}
	\boxed{
		\Lambda(P)=G'(P).
	}
\end{equation}
This identity relates the propagation of a small disagreement to the sensitivity of the recursive map.
At the endpoint of the Bethe-lattice hysteresis discontinuity, the critical fixed point satisfies
\begin{equation}
	P_\star=G(P_\star),
	\qquad
	G'(P_\star)=1,
	\qquad
	G''(P_\star)=0.
\end{equation}
The first two relations, together with $\Lambda(P)=G'(P)$, imply
\begin{equation}
	\boxed{\Lambda(P_\star)=1.}
\end{equation}
Thus, at the Bethe-lattice hysteresis critical point, the linearized propagation of a microscopic disagreement is marginal.
This benchmark shows that the Bethe-lattice avalanche condition is a local marginality condition, conceptually distinct from the finite-dimensional, system-size noise-sensitivity criterion developed in the Letter.

\section{Relation to RFIM disorder-chaos studies}
Previous RFIM work has studied the sensitivity of equilibrium ground-state spin overlap to quenched-disorder perturbations, a phenomenon often referred to as disorder chaos \cite{SAlavaRieger1998}.
That problem differs from the present one in both dynamics and observable: here we study a nonequilibrium hysteretic terminal event and ask for the order of microscopic information required to predict it.
We therefore do not interpret our result as the discovery of disorder chaos in the RFIM.
Rather, our contribution is to combine the established KMS influence bounds with a Gaussian Russo formula and express the resulting criterion through the switching-field distribution, thereby connecting a measurable macroscopic response to terminal-outcome information order.


\begin{thebibliography}{999}
\bibitem{BenjaminiKalaiSchramm1999} I. Benjamini, G. Kalai, and O. Schramm, Noise sensitivity of Boolean functions and applications to percolation, Publ. Math. IH\'ES {\bf 90}, 5 (1999).

\bibitem{SchrammSteif2010} O. Schramm and J. E. Steif, Quantitative noise sensitivity and exceptional times for percolation, Ann. Math. {\bf 171}, 619 (2010).

\bibitem{GarbanPeteSchramm2010} C. Garban, G. Pete, and O. Schramm, The Fourier spectrum of critical percolation, Acta Math. {\bf 205}, 19 (2010).

\bibitem{FriedgutKalai1996} E. Friedgut and G. Kalai, Every monotone graph property has a sharp threshold, Proc. Am. Math. Soc. {\bf 124}, 2993 (1996).

\bibitem{KellerMosselSen2012} N. Keller, E. Mossel, and A. Sen, Geometric influences, Ann. Probab. {\bf 40}, 1135 (2012).

\bibitem{KellerMosselSen2014} N. Keller, E. Mossel, and A. Sen, Geometric influences II: Correlation inequalities and noise sensitivity, Ann. Inst. Henri Poincar\'e Probab. Stat. {\bf 50}, 1121 (2014).

\bibitem{SethnaEtAl1993} J. P. Sethna, K. Dahmen, S. Kartha, J. A. Krumhansl, B. W. Roberts, and J. D. Shore, Hysteresis and hierarchies: Dynamics of disorder-driven first-order phase transformations, Phys. Rev. Lett. {\bf 70}, 3347 (1993).

\bibitem{DahmenSethna1996} K. Dahmen and J. P. Sethna, Hysteresis, avalanches, and disorder-induced critical scaling: A renormalization-group approach, Phys. Rev. B {\bf 53}, 14872 (1996).

\bibitem{PerkovicDahmenSethna1999} O. Perkovi{\'c}, K. A. Dahmen, and J. P. Sethna, Disorder-induced critical phenomena in hysteresis: Numerical scaling in three and higher dimensions, Phys. Rev. B {\bf 59}, 6106 (1999).

\bibitem{AlavaRieger1998} M. Alava and H. Rieger, Chaos in the random field Ising model, Phys. Rev. E {\bf 58}, 4284 (1998).

\bibitem{PachecoEtAl2009} J. M. Pacheco, F. C. Santos, M. O. Souza, and B. Skyrms, Evolutionary dynamics of collective action in $N$-person Stag Hunt dilemmas, Proc. R. Soc. B {\bf 276}, 315 (2009).

\bibitem{Watts2002} D. J. Watts, A simple model of global cascades on random networks, Proc. Natl. Acad. Sci. U.S.A. {\bf 99}, 5766 (2002).

\bibitem{Blume1995} L. E. Blume, The statistical mechanics of best-response strategy revision, Games Econ. Behav. {\bf 11}, 111 (1995).

\bibitem{MondererShapley1996} D. Monderer and L. S. Shapley, Potential games, Games Econ. Behav. {\bf 14}, 124 (1996).

\bibitem{KindlerKirshnerODonnell2018} G. Kindler, N. Kirshner, and R. O'Donnell, Gaussian noise sensitivity and Fourier tails, Isr. J. Math. {\bf 225}, 71 (2018).

\bibitem{Supplement} See Supplemental Material for derivations, model and simulation details, robustness tests, and finite-size statistics.

\bibitem{DharShuklaSethna1997} D. Dhar, P. Shukla, and J. P. Sethna, Zero-temperature hysteresis in the random-field Ising model on a Bethe lattice, J. Phys. A: Math. Gen. {\bf 30}, 5259 (1997).

\bibitem{KuntzEtAl1999} M. C. Kuntz, O. Perkovi{\'c}, K. A. Dahmen, B. W. Roberts, and J. P. Sethna, Hysteresis, avalanches, and noise, Comput. Sci. Eng. {\bf 1}, 73 (1999).

\bibitem{ZimmaroEtAl2024} F. Zimmaro, S. Galam, and M. A. Javarone, Asymmetric games on networks: Mapping to Ising models and bounded rationality, Chaos Solitons Fractals {\bf 181}, 114666 (2024).
\end{thebibliography}

\begin{thebibliography}{99}
	
	\bibitem{SKellerMosselSen2012} N. Keller, E. Mossel, and A. Sen, Geometric influences, Ann. Probab. {\bf 40}, 1135 (2012).
	
	\bibitem{SKellerMosselSen2014} N. Keller, E. Mossel, and A. Sen, Geometric influences II: Correlation inequalities and noise sensitivity, Ann. Inst. Henri Poincar\'e Probab. Stat. {\bf 50}, 1121 (2014).
	
	\bibitem{SPachecoEtAl2009} J. M. Pacheco, F. C. Santos, M. O. Souza, and B. Skyrms, Evolutionary dynamics of collective action in $N$-person Stag Hunt dilemmas, Proc. R. Soc. B {\bf 276}, 315 (2009).
	
	\bibitem{SWatts2002} D. J. Watts, A simple model of global cascades on random networks, Proc. Natl. Acad. Sci. U.S.A. {\bf 99}, 5766 (2002).
	
	\bibitem{SZimmaroEtAl2024} F. Zimmaro, S. Galam, and M. A. Javarone, Asymmetric games on networks: Mapping to Ising models and bounded rationality, Chaos Solitons Fractals {\bf 181}, 114666 (2024).
	
	\bibitem{SBlume1995} L. E. Blume, The statistical mechanics of best-response strategy revision, Games Econ. Behav. {\bf 11}, 111 (1995).
	
	\bibitem{SMondererShapley1996} D. Monderer and L. S. Shapley, Potential games, Games Econ. Behav. {\bf 14}, 124 (1996).
	
	\bibitem{SDharShuklaSethna1997} D. Dhar, P. Shukla, and J. P. Sethna, Zero-temperature hysteresis in the random-field Ising model on a Bethe lattice, J. Phys. A: Math. Gen. {\bf 30}, 5259 (1997).
	
	\bibitem{SAlavaRieger1998} M. Alava and H. Rieger, Chaos in the random field Ising model, Phys. Rev. E {\bf 58}, 4284 (1998).
	
\end{thebibliography}
\end{document}